\def\beq{\begin{equation}}
\def\eeq{\end{equation}}
\def\beeq{\begin{eqnarray}}
\def\beeqn{\begin{eqnarray*}}
\def\eeeq{\end{eqnarray}}
\def\eeeqn{\end{eqnarray*}}
\def\a{\alpha}

\def\l{\lambda}                 
\def\m{\mu}
\def\n{\nu}

\def\r{\rho}
\def\s{\sigma}                  
\def\th{\theta}

\def\der{\partial}

\newcommand{\lp}{\left(}
\newcommand{\rp}{\right)}

\newcommand{\alp}{\alpha '}

\newcommand{\no}{\nonumber}

\def\tr{\,\mbox{Tr}\,}
\def\frac#1#2{ {{#1} \over {#2} }}

\documentclass[letterpaper]{JHEP3} 

\usepackage{epsfig,multicol}
\newcommand\fverb{\setbox\pippobox=\hbox\bgroup\verb}
\newcommand\fverbdo{\egroup\medskip\noindent%
			\fbox{\unhbox\pippobox}\ }
\newcommand\fverbit{\egroup\item[\fbox{\unhbox\pippobox}]}
\newbox\pippobox
\title{One-loop unitarity of string theories in a constant external background 
and their Seiberg-Witten limit}
\author{A. Bassetto, A. Torrielli\\
Dipartimento di Fisica ``G.Galilei", Via Marzolo 8, 35131
Padova, Italy\\
INFN, Sezione di Padova, Italy\\
E-mail: \email{antonio.bassetto@pd.infn.it},\email{alessandro.torrielli@pd.infn.it}}
\author{R. Valandro\\
International School for Advanced Studies, Via Beirut 2, 34100 Trieste, Italy\\
E-mail: \email{valandro@sissa.it}}
\preprint{DFPD 03/TH 44, SISSA 98/2003/EP}
\received{} 		%%
\revised{}
\accepted{}		%% These are for published papers.

\abstract{Perturbative spectra and related factorization properties of one-loop
open string amplitudes in the presence of a constant external background $B$ are 
analysed in detail. While the pattern of the closed string spectrum, obtained 
after a careful
study of the properly symmetrized amplitudes, turns out to be unaffected 
by the presence
of $B$, a series of double open-string poles, which would be absent when 
$B$ is turned off,
can couple owing to a partial symmetry loss. These features are studied first in a 
bosonic setting and then generalized in the more satisfactory superstring context.
When the background is of an ``electric'' type, a classical perturbative
instability is produced beyond a critical value of the electric field. In the
Seiberg-Witten limit this instability is the origin of the unphysical tachyonic
cut occurring in the non-planar amplitudes of the corresponding noncommutative
field theories.}

\begin{document}

%\noindent

\dedicated{}

%\begin{document} 

%\maketitle  IS IGNORED %%%%%%%%%%%
\section{Introduction}

String theories in the presence of a constant background field ($B$-field)
have been often considered in the recent literature; in particular Seiberg
and Witten \cite{sw} have shown they give rise in a suitable limit
to field theories defined on a space of noncommuting variables (NCFT).

In turn NCFT may suffer from the lack of covariance and, when the $B$-field
is of electric type, their spectrum has severe difficulties with respect
to perturbative S-matrix unitarity \cite{gm,agm,bgnv}. 
Usual analytic properties of amplitudes are
lost as well, as a consequence of non-locality.

Many of these unpleasant features are not shared by the parent string
theory \cite{sst2,barbrab}; therefore it looks interesting to re-examine the spectrum
of the latter as a first step in order to point out where possible 
differences are generated.
Such an analysis does not seem to have been carried out in full 
detail, at least to our knowledge, in spite of the huge amount
of literature on the subject (see for instance \cite{ad,KieLee,gkmrs,liumich1,crs,
gmms,abz,arm,rey}).

A first approach to this problem was undertaken in ref.\cite{mio},
where the singularities of the bosonic two-point amplitudes were
studied, after a suitable off-shell continuation \cite{bcr}, in 
order to perform a comparison with the ones occurring in the
corresponding noncommutative field  amplitudes.
In particular the appearance in these amplitudes of an unphysical cut 
when the noncommutativity parameter involves the time variable
was related to the classical instability of the corresponding 
string theory in the presence of a background external field
of an ``electric'' type.

The analysis above was generalized in \cite{roberto,mio1} to the four-point
``on-shell'' tachyonic amplitude, thereby avoiding any possible
trouble with off-shell continuations.

In the following we start considering again the bosonic string case,
namely we study four-string amplitudes in the presence of a background field, 
first at tree level and
then at one loop level, paying particular attention at the spectrum
of the various singularities and at the related factorization properties. 
Later we repeat the above analysis for superstring amplitudes,
where unsatisfactory unphysical features are no longer present. 

We consider the bosonic case first since, although affected by the
well-known pathologies, it still exhibits in a simpler setting
most of the features we shall later encounter in the better
grounded superstring context. 

The motivation for looking at tree
level amplitudes is prompted by the desire of exploring the
open string spectrum in the presence of the $B$-field; such a field
entails the occurrence of extra poles which otherwise would decouple for
spatial symmetry reasons. One-loop amplitudes are instead
essential to capture effects from the closed string spectrum,
in particular the open-closed string vertices which can be
obtained via factorization. These are the amplitudes related
to the non-planar ones in the NCFT limit.

In the last part of our work such results will be extended
to the superstring case, where pathologies related to ghosts
and to redundant dimensions are absent. Here a stack of $N$ $Dp$-branes
is to be introduced in order to deal with the $U(N)$ group \cite{tasi,lm} 
and the $B$-field
is chosen with non-vanishing components only in the directions of
the branes. Actually two stacks of branes will be considered
at a relative distance $\vec{Y}$, with string exchanges in between them.
Again the presence of the $B$-field will reduce part of existing symmetries
and thereby allow the presence of further poles which would
decouple otherwise. 
Consistency with factorization properties
following from unitarity will be checked explicitly. In particular the two
different situations, characterized by a field $B$ of a ``magnetic'' and of
an ``electric'' type respectively, are discussed and compared. In the first case
the theory does not exhibit any perturbative instability, at least in the
one-loop amplitude we have considered. In the electric case instead
a classical instability
appears in the perturbative string amplitude when the ``electric'' field 
overcomes a critical
value \cite{b,bp,sst1}, due to an uncontrolled growth of the oscillation of modes in
the direction parallel to the field. Corresponding to this value a violation
of unitarity occurs in the form of a cut in the
complex squared energy plane of tachyonic type.

When the Seiberg-Witten limit is considered leading to an effective
field theory of a noncommutative type, the electric field is necessarily
pushed into the instability region and the resulting theory is sick.
 
In sect. 2 the bosonic case is studied, first at the tree level, pointing
out the peculiarities related to the presence of the $B$-field, and then
at the one loop where the interplay between closed and open sectors becomes
apparent. Sect.3 is devoted to superstring amplitude, where analogous features
occur in the scattering of strings lying on two stacks of branes at a relative
distance $\vec{Y}$. When eventually the Seiberg-Witten limit is
performed, several peculiarities of noncommutative field theories find their
{\it raison d'\^etre} in the corresponding features of the parent string theory.
Final comments are the content of sect.4 .

\section{The bosonic case}

\subsection{Tree amplitudes}

Before entering {\it in medias res},
some notations and definitions have to be recalled. The following expression
\beq \label{ac}
S_{bos}={{1}\over{4\pi {\alpha}'}}\int_{C}d^2 z \, (g_{\mu \nu} 
{\partial}_a X^{\mu} {\partial}^a X^{\nu} - 
2i \pi {\alpha}' B_{ij} {\epsilon}^{ab}{\partial}_a X^{i} 
{\partial}_b X^{j})     
\eeq
is the action of a bosonic open string attached to a $D$-brane lying in the 
first $p+1$ dimensions and coupled to an antisymmetric constant background
\cite{sw,ad}.
We denote by latin letters $i,j,...$ the components along the brane. 
The open string parameters are:
\beq \label{openG}
G=(g-2\pi {\alpha}' B)g^{-1} (g+2\pi {\alpha}' B),
\eeq
\beq \label{thet}
\theta=-{(2\pi{\alpha}' )}^2 {(g+2\pi {\alpha}' B)}^{-1}B {(g-2\pi {\alpha}' B)}^{-1}.
\eeq
Here $G_{ij}$ and $g_{ij}$ are the open and the closed-string metric tensors,
respectively.

We consider for simplicity the four-tachyon tree amplitude and
introduce the usual Mandelstam variables $s=-(k_1+k_2)^2, t=-(k_1+k_4)^2$ and 
$u=-(k_1+k_3)^2$ using the metric tensor $G_{ij}$. The
tachyons are on shell, $k_i^2=-m^2,\quad i=1,...,4.$. Our open string
metric is $(-1,1,...,1)$
and we choose our units so that $m^2=-2, (\alpha'=1/2)$. As a consequence $s+t+u=-8$.

The presence of the $B$-field affects the 
familiar Veneziano expression by a phase factor
\beeq \label{vene}
&&A(k_1,...,k_4)= \frac{\Gamma(-1-s/2) \Gamma(-1-t/2)}{\Gamma(-2-s/2-t/2)}
\exp\Big(\frac{-i}2 (k_2\tilde k_1+k_3\tilde k_1+k_3\tilde k_2)\Big) \nonumber\\
&+& \,\mbox{non-cyclic permutations},
\eeeq 
where $(\tilde k)^i=\theta^{ij} (k)_j$ and $\theta^{ij}=\pi \Big(\frac 1{g+\pi B}\Big)
^{ij}_A$. 
$(\quad)_A$ denotes the antisymmetric part of a matrix.

No Chan-Paton factors are considered for the time being; we restrict
ourselves to the $U(1)$ case. This is not a real limitation and will
be removed when considering one-loop amplitudes. 

We are now interested in studying the pole of eq.(\ref{vene}) at $s=0$. One can
easily realize that the relevant terms are
\beeq \label{vener}
&&A(k_1,...,k_4)|_{rel}= \frac{\Gamma(-1-s/2) \Gamma(-1-t/2)}{\Gamma(-2-s/2-t/2)}
\exp\Big(\frac{-i}2 (k_2\tilde k_1+k_3\tilde k_1+k_3\tilde k_2)\Big) \nonumber\\
&+& (k_1 \leftrightarrow k_2)+ (k_3 \leftrightarrow k_4).
\eeeq 
The amplitude on the pole behaves like
\beeq \label{venep}
&&A(k_1,...,k_4)\simeq -\frac{2}{s} \exp \Big(\frac{-i}2 k_3(\tilde k_1 +\tilde k_2)\Big)
\Big[(2+t/2)\exp (\frac{-i}2 k_2\tilde k_1)+(2+u/2)\exp (\frac{i}2 k_2\tilde k_1)\Big]
\nonumber \\
&&+(k_3\leftrightarrow k_4)= \frac{2}{s}(t-u) \sin \frac{k_1\tilde k_2}2
\sin \frac{k_3\tilde k_4}2,
\eeeq
where momentum conservation and the mass shell conditions have been taken
into account.

It is immediately clear that, in the absence of the $B$-field ($\theta^{ij}=0$),
the above residue vanishes: The two tachyons cannot couple to a photon while
respecting Bose statistics. In the presence of the $B$-field a nice factorization
occurs of the residue as the product of two vertices, each carrying a Moyal
phase, as expected from the result in ref.\cite{sw}. 
Indeed the vertex tachyon-tachyon-photon takes the form
\beq \label{ver}
V^{i}(k_1,k_2)\simeq (k_1^{i}-k_2^{i}) \sin \frac{k_1\tilde k_2}2,
\eeq
and the presence of the Moyal phase is crucial to comply with Bose statistics.
We notice that the transversality condition $(k_1+k_2)\cdot V=0$ is satisfied.

\subsection{One-loop amplitudes}

We turn now our attention to the study of one-loop amplitudes, always in
the presence of a constant field $B$. In so doing
our purpose is to explore the features of closed string poles in
addition to the open string ones.

At one loop the string world-sheet is the cylinder $C_2 = 
\{0\leq \Re w\leq 1, w = w + 2 i \tau$\}.

The one loop propagator with the boundary conditions imposed by the $B$-term,
%\footnote{The reader is referred to the literature \cite{bcr,gkmrs,kl,l,crs}}, 
can be found in \cite{ad}. If one sets $w=x+iy$, the relevant propagator 
on the boundary of the cylinder ($x=0,1$) can be written as 
\beq
\label{prp}
G(y,y')={{1}\over{2}}{\alpha}' g^{-1} \log q - 2 {\alpha}' G^{-1} \log 
\Big[{{q^{{1}\over {4}}}\over {D(\tau)}} \, {\vartheta_4}({{|y-y'|}
\over {2 \tau}}, {{i}\over {\tau}})\Big],\, \, \, x\neq x', 
\eeq
\beq \label{prp=}
G(y,y')={{\pm i \theta }\over{2}} {\epsilon}_{\perp} (y-y') - 
2{\alpha}' G^{-1} \log \Big[{{1}\over {D(\tau)}} {\vartheta_1}
({{|y-y'|}\over {2 \tau}}, {{i}\over {\tau}})\Big],\, \, \, x=x',
\eeq  
where $q=e^{-{{\pi}\over{\tau}}}$, $\pm$ correspond to $x=1$ 
and $x=0$ respectively, and ${\epsilon}_{\perp} (y) = sign(y) - {{y}\over{\tau}}$.

Here ${\vartheta_{1,4}}(\nu, \tau )$ are Jacobi theta functions, while 
$D(\tau)={\tau}^{-1} 
{[\eta({i\over \tau})]}^3$ and $\eta$ is the Dedekind eta function \cite{p}.

With this propagator and the suitable modular measure, the amplitude for the 
insertion of $N$ tachyonic vertex operators at $x=1$ and $M-N$ at $x=0$ turns 
out to be \cite{ad}: 
\beeq
\label{ampl}
A_{1,\cdots, M}&=&{\cal{N}}_0 \int_0^{\infty} {{d\tau}\over{\tau}} 
{\tau}^{-{{d}\over{2}}} {[\eta(i\tau)]}^{2-d} q^{{{1}\over{2}}
{\alpha}' Kg^{-1} K}  \nonumber \\
&&\times \int [dy] \, \prod_{i=1}^N \prod_{j=N+1}^M 
{\Bigg[ q^{{1}\over {4}} \, {\vartheta_4}({{|y_i - y_j|}
\over {2 \tau}}, {{i}\over {\tau}}) / D(\tau)\Bigg]}^
{2 {\alpha}' k_i G^{-1} k_j } \no \\
&&\times \prod_{1=i<j}^N e^{- {{1}\over{2}}i {\epsilon}_{\perp} 
(y_i - y_j)k_i \theta k_j} {\Bigg[ {\vartheta_1}({{|y_i -y_j|}
\over {2 \tau}}, {{i}\over {\tau}}) / D(\tau)\ \Bigg]}^
{2 {\alpha}' k_i G^{-1} k_j }\nonumber \\
&&\times \prod_{N+1=i<j}^M e^{ {{1}\over{2}}i {\epsilon}_{\perp} 
(y_i - y_j)k_i \theta k_j} {\Bigg[ {\vartheta_1}({{|y_i -y_j|}
\over {2 \tau}}, {{i}\over {\tau}}) / D(\tau)\ \Bigg]}^{2 {\alpha}' k_i G^{-1} k_j }.
\eeeq
Here ${\cal{N}}_0$ is a normalization constant, $d=p+1$, and $K=\sum_{i=1}^N k_i$ 
is the sum of all momenta associated with the vertex operators inserted on the 
$x=1$ boundary. The integration region $[dy]$ for the variables $y_i$ will be 
specified later on. 

Whenever $N\neq0$ and $M>N$, this amplitude corresponds to non-planar graphs, 
the traces of the relevant Chan-Paton matrices being understood.

For electric backgrounds it is well known that problems arise when the electric field 
approaches a critical value $E_{cr}$. Beyond it a classical instability 
occurs \cite{b,bp,sst1} both for neutral (which is the case we consider here) 
and for charged open strings, related to an uncontrolled growth of the oscillation 
amplitude of modes in the direction parallel to the field. 

This phenomenon coexists with the quantum instability of purely charged strings 
due to pair production in any electric field (even in a sub-critical one), 
which is the analog of the Schwinger phenomenon in particle electrodynamics.
However, at variance with the latter, it has no analog in particle field theory
\footnote{In passing, we recall that for neutral strings ($q$-charge on one end, 
$-q$ on the other), one has $|E_{cr}|=1/(2\pi \alpha' |q|)$. 
For charges $q_1\neq q_2$ on the two boundaries, one finds that the pair 
production rate diverges at a critical value $E_{cr} = 1/(2\pi \alpha' |max\, q_i |)$}. 
We will discuss it in subsect 3.2 .

\subsection{Closed string poles}

We write the one-loop non-planar amplitude for a scattering of four bosonic open 
string tachyons in $d=26$ ($D25$-brane) in the presence of a constant antisymmetric 
background as follows:
\beq
\label{calA}
{\cal{A}} = A(1,2,3,4) \tr [\lambda_1 \lambda_2 ] \tr [\lambda_3 \lambda_4 ]
+ \mbox{non-trivial permutations}, 
\eeq
where the amplitude $A(1,2,3,4)$ above, after specializing eq.(\ref{ampl}) 
(see Appendix A),  takes the form 

\begin{figure}
\label{figura1}
\centerline{\includegraphics[width=13.5cm]{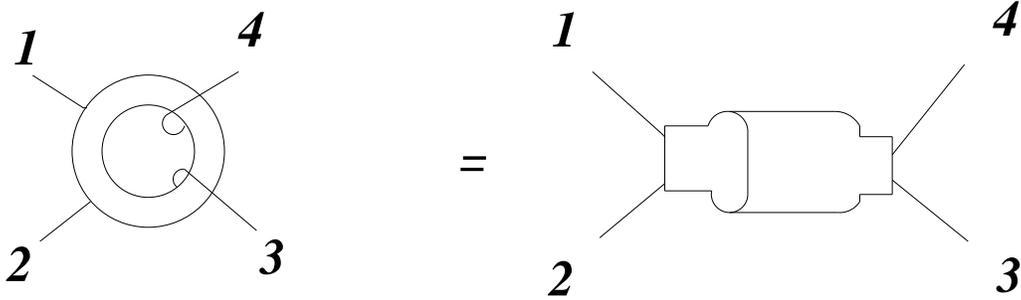}}
\caption{Diagram representing the amplitude $A(1,2,3,4)$}
\end{figure}

\beeq
\label{A1234}
&&A= {\cal{N}} \int_0^1 {{dq}\over{q^3}} {[f(q^2 )]}^{- 16} 
q^{{{1}\over {4}}K g^{- 1} K} \int_0^1 d\nu_1 \int_0^{\nu_1} d\nu_2 
\int_0^{\nu_2} d\nu_3 e^{- {{i}\over {2}} [k_1 \theta k_2 (1 - 2 \nu_{12})- k_3 
\theta k_4 (1 - 2 \nu_{3})]} 
\nonumber\\
&&{\left[[\sin \pi \nu_{12} \prod_1^{\infty} \left( 1 - 2q^{2n}\cos 2\pi 
\nu_{12} + q^{4n}\right)]\,\,[\sin \pi \nu_{3} 
\prod_1^{\infty} \left( 1 - 2q^{2n}\cos 2\pi \nu_{3} + q^{4n}\right)]\right]}
^{- 2 - (s/2)} \\
&&{\left[\prod_1^{\infty} \left( 1 - 2q^{2n-1}\cos 2\pi \nu_{13} + q^{4n-2}
\right)\right]}^{- 2 - (u/2)} {\left[\prod_1^{\infty} \left( 1 - 2q^{2n-1}
\cos 2\pi \nu_{1} + q^{4n-2}\right)\right]}^{- 2 - (t/2)} \nonumber \\   
&&{\left[\prod_1^{\infty} \left( 1 - 2q^{2n-1}\cos 2\pi \nu_{23} 
+ q^{4n-2}\right)\right]}^{- 2 - (t/2)} {\left[\prod_1^{\infty} 
\left( 1 - 2q^{2n-1}\cos 2\pi \nu_{2} + q^{4n-2}\right)\right]}^{- 2 - (u/2)}.
\nonumber
\eeeq
Eq.(\ref{A1234}) is the amplitude for a fixed order of the external momenta 
(see the diagram in fig. 1); the coordinates 
$\nu_i$ of the vertex operator insertions are taken to be cyclically ordered from zero 
to one on each of the two boundaries, with $\nu_4$ fixed to zero 
in order to remove the residual gauge. All possible non-trivial 
permutations of the labels of the external particles are to be summed over. 
The ones concerning $k_3$ are accounted for by extending the region
of integration over $\nu_3$ from $0$ to $1$. Henceforth we shall denote
by the symbol $A(1,2)$ such an extension ($0\le \nu_3\le 1)$. 
We have expressed the amplitude 
using the variable $q = \exp [- \pi / \tau ]$, which is the traditional one that 
reference \cite{gsw} uses for the spectral analysis in the case without $B$-field, 
in order to have a direct comparison. External momenta are taken to be on the 
tachyon mass-shell.
Indices are raised and lowered with the open string metric $G^{-1}$. 
We define $\nu_{rs}=\nu_r - \nu_s$; $K = k_1 + k_2$ is the total momentum  
entering through one boundary and leaving from the other. We also define 
$f(x) = \prod_{n=1}^{\infty} (1 - x^n)$ as in \cite{gsw}.

It is well known that, in the case without the $B$-field, this amplitude 
presents a rich pattern of singularities, each of them being required by  
unitarity, corresponding to precise intermediate exchanges. 

First of all, if one takes the amplitude $A(1,2)$ as an example, and analyzes 
it in the variable $s$, setting $\theta=0$ and $g=G$ inside the integral,
one finds a series of simple poles when $s$ equals the masses of the closed 
string tower. This is required by unitarity, because a diagram like the one of fig.1 
is topologically equivalent to a tree-level closed string exchange between 
the sets of particles $\{1,2\}$ and $\{3,4\}$. 

These poles emerge when one explores the small region of the integration
over the $q$-variable; in particular, one easily sees that the amplitude 
can be written as follows
\beq
\label{sommapoli}
A(1,2) = \sum_{n=0}^{\infty} \int_0^1 dq \, q^{- 3 + n - (s / 4)} \, \, {a_n}(s,t,u),
\eeq
namely
\beq
\label{sommapoli2}
A(1,2) = \sum_{n=0}^{\infty} {{1}\over{(- 2 + n - (s / 4))}} \, {a_n}(s,t,u).
\eeq
>From this expansion one might naively conclude that poles occur at the masses 
$s = 4n - 8 = -8, -4, 0, \cdots$, which, unless canceled by the
sum over permutations, would be at odds with unitarity, since the expected closed 
string spectrum is $s = -8, 0, \cdots$.
Actually quantities like $a_1, a_3,...$ cancel on their own, as one can check 
explicitly by 
performing the integrals over the $\nu_i$ or as a particular case of Appendix B. 

The first thing to be checked is whether such a cancellation persists in the
presence of the $B$-field; we exhibit it here explicitly for the lowest level $n=1$.
The remarkable proof of cancellation for higher odd values of $n$ is 
given in Appendix B.
It is easy to realize from eq.(\ref{A1234}) that,
even in the presence of the $B$-field, a formula analogous to (\ref{sommapoli2}) 
can be written
\beq
\label{sommapoli2theta}
A(1,2) = \sum_{n=0}^{\infty} {{1}\over{(- 2 + n - (s_{CL} / 4))}} \, 
{\alpha_n}(s,t,u, k_1 \theta k_2 , k_3 \theta k_4 ).
\eeq
We have defined $s_{CL}=-K g^{-1} K$ the squared energy variable in the closed channel, 
which is different from the one in the open channel $s$ as long as $B$ is different 
from zero. 

We perform a Taylor expansion of the integrand in the expression of $A$ 
up the first order in $q$ and get
\beeq
\label{alfa1}
\alpha_1 &=& {\cal{N}} \int_0^1 d\nu_1 \int_0^{\nu_1} d\nu_2 
\int_0^1 d\nu_3 e^{- {{i}\over {2}} k_1 \theta k_2 (1 - 2 \nu_{12} )} 
e^{{{i}\over {2}} k_3 \theta k_4 (1 - 2 \nu_{3} )} {\left( \sin \pi \nu_{12} 
\sin \pi \nu_3 \right)}^{- 2 - (s/2)}\nonumber \\
&&\times \left[ (4 + u) (\cos 2 \pi \nu_{13} + \cos 2 \pi \nu_2 ) 
+ (4 + t) (\cos 2 \pi \nu_1 + \cos 2 \pi \nu_{23} )\right].      
\eeeq
We now change variables to $p=\nu_{12}$; the integration region 
becomes $\int_0^1 dp \int_p^1 d\nu_1 \int_0^1 d\nu_3 $. We easily integrate 
over $\nu_1$ and, after use of trigonometric identities, obtain
\beeq
\label{alfa1p}
&&\alpha_1 = \frac{- 4 {\cal N}}{2\pi} \int_0^1 dp \int_0^1 d\nu_3  
e^{- {{i}\over {2}} k_1 
\theta k_2 (1 - 2 p )} e^{{{i}\over {2}} k_3 \theta k_4 (1 - 2 \nu_{3} )} 
{\left( \sin \pi p \sin \pi \nu_3 \right)}^{- 2 - (s/2)} \\
&&\times \left[ (8 +t + u) (\sin \pi p \cos \pi p - \sin \pi p \cos \pi p \, 
{\sin}^2 \pi \nu_3 ) + (u - t) \sin \pi \nu_3 \cos \pi \nu_3 \, 
{\sin}^2 \pi p \right] \nonumber .      
\eeeq
Using the formulas (\ref{a1}), we finally get
\beeq
\label{alfa1finale}
\alpha_1 &=& {{{2^{s} \, i \, \cal{N}}}\over {\pi^2 }}  \Bigg[ {{4 (8 + t + u) 
\, \Gamma (-1 - {{s}\over{2}}) \Gamma(- {{s}\over{2}}) \, k_1 \theta k_2}\over 
{\Gamma (- {{s}\over{4}} + {{k_3 \theta k_4 }\over {2 \pi}}) \Gamma (- {{s}\over{4}} 
- {{k_3 \theta k_4 }\over {2 \pi}}) \Gamma (1 - {{s}\over{4}} 
+ {{k_1 \theta k_2 }\over {2 \pi}}) \Gamma (1 - {{s}\over{4}} - {{k_1 \theta k_2 }
\over {2 \pi}})}}\nonumber \\
&& - {{(8 + t + u) \, \Gamma (1 - {{s}\over{2}}) \Gamma(- {{s}\over{2}}) 
\, k_1 \theta k_2}\over {\Gamma (1 - {{s}\over{4}} + {{k_3 \theta k_4 }\over {2 \pi}}) 
\Gamma (1 - {{s}\over{4}} - {{k_3 \theta k_4 }\over {2 \pi}}) \Gamma (1 - {{s}\over{4}} 
+ {{k_1 \theta k_2 }\over {2 \pi}}) \Gamma (1 - {{s}\over{4}} - {{k_1 \theta k_2 }
\over {2 \pi}})}}\nonumber \\
&& - {{(u - t) \, \Gamma (1 - {{s}\over{2}}) \Gamma(- {{s}\over{2}}) \, 
k_3 \theta k_4}\over {\Gamma (1 - {{s}\over{4}} + {{k_3 \theta k_4 }\over {2 \pi}}) 
\Gamma (1 - {{s}\over{4}} - {{k_3 \theta k_4 }\over {2 \pi}}) \Gamma (1 - {{s}\over{4}} 
+ {{k_1 \theta k_2 }\over {2 \pi}}) \Gamma (1 - {{s}\over{4}} - {{k_1 \theta k_2 }
\over {2 \pi}})}}\Bigg].
\eeeq
The first thing to notice is that, in the limit $\theta\to 0$, all this expression 
vanishes and we recover the mentioned result when $B=0$. On the contrary, 
for a non-vanishing $\theta$, (\ref{alfa1finale}) is different from zero. 
As already mentioned, this would contradict unitarity, since the pattern of the
closed string 
spectrum is unaffected by the presence of $B$. 
Now the cancellation is subtler; we notice that the expression 
(\ref{alfa1finale}) is odd 
under the exchange $k_1\leftrightarrow k_2$, as $\theta$ is antisymmetric and 
$t\leftrightarrow u$ under such an exchange (the Chan-Paton factors in 
(\ref{A1234}) remain the same). 
Therefore, the sum over the other three diagrams considered in (\ref{calA}) 
gets rid of such an unwanted
singularity.

The closed poles in the non-planar diagram 
of fig.1 correspond to the traditional pattern.
To get the cancellation of the unwanted closed poles one needs to sum over permutations: 
The expected structure is recovered only at the level of the complete four-point 
amplitude.
\footnote{This could also be interpreted as an indication of the need to sum over 
all possible configurations in dealing with amplitudes in the presence of the 
$B$-background.}

\subsection{Open string poles}

The amplitude $A(1,2,3,4)$ in the case of $B=0$ (or, equivalently, $\theta=0$) presents 
as well a series of open string poles in the s-channel with singlet quantum numbers. 
They arise when the $\nu_i$'s on the same boundary of the diagram  
in fig. 1 get close to a common value. 

When $B \neq 0$, $s_{CL}\neq s$ and there are two variables in which singularities
may occur. To simplify our analysis, we single out the first non-vanishing residue
in the variable $s_{CL}$, $\alpha_0$, and examine its behaviour as a function
of $s$. The very same Taylor expansion of the integrand of $A(1,2)$
in the variable $q$ reveals that $\alpha_0$ is given by
\beq
\label{alfa0p}
\alpha_0 = {\cal{N}} \int_0^1 dp \int_0^1 d\nu_3 e^{- {{i}\over {2}} k_1 
\theta k_2 (2 p - 1)} e^{{{i}\over {2}} k_3 \theta k_4 (1 - 2 \nu_{3} )} \, p 
\, {\left( \sin \pi p \sin \pi \nu_3 \right)}^{- 2 - (s/2)}
\eeq
and, using formulas easily derived from (\ref{a1}), we find
\beq
\label{alfa0tot}
\alpha_0 = {{- i {\cal{N}} 2^{3+s} \, \, \Gamma^2 (- 1 - {{s}\over {2}}) 
\left[ \pi i + \psi (- {{s}\over {4}} + {{k_1 \theta k_2 }\over {2 \pi}}) 
- \psi (- {{s}\over {4}} - {{k_1 \theta k_2 }\over {2 \pi}})\right]}
\over {\pi \Gamma (- {{s}\over {4}} + {{k_3 \theta k_4 }\over {2 \pi}}) 
\Gamma (- {{s}\over {4}} - {{k_3 \theta k_4 }\over {2 \pi}}) 
\Gamma (- {{s}\over {4}} + {{k_1 \theta k_2 }\over {2 \pi}}) 
\Gamma (- {{s}\over {4}} - {{k_1 \theta k_2 }\over {2 \pi}})}},
\eeq
with $\psi (z) = \Gamma' (z) / \Gamma(z)$.

As a first remark, we notice that, when $\theta$ goes to zero, 
this result smoothly tends to the expression  
\beq
\label{alfa0theta0}
a_0 = {{{\cal{N}} \, \Gamma^2 ( - {{2+s}\over {4}})}\over 
{2 \pi \Gamma^2 (- {{s}\over {4}})}}
\eeq
of eq.(\ref{sommapoli2}), which exhibits double poles at $s = -2, 2, 6, \cdots$.

After the symmetrization $k_1\leftrightarrow k_2$, the terms containing the 
$\psi$-function cancel and we are left with
\beq
\label{alfa0totperm}
\alpha_0 (1,2) + \alpha_0 (2,1)= {{{\cal{N}} 2^{4+s} \, \, 
\Gamma^2 (- 1 - {{s}\over {2}})}\over {\Gamma (- {{s}\over {4}} 
+ {{k_3 \theta k_4 }\over {2 \pi}}) \Gamma (- {{s}\over {4}} 
- {{k_3 \theta k_4 }\over {2 \pi}}) \Gamma (- {{s}\over {4}} 
+ {{k_1 \theta k_2 }\over {2 \pi}}) \Gamma (- {{s}\over {4}} 
- {{k_1 \theta k_2 }\over {2 \pi}})}}.
\eeq

Now the poles, which were missing when $\theta=0$, are switched on, the presence of 
the $B$-field providing extra structures. Indeed the gamma function in the numerator 
has poles at $s = -2, 0, 2, 4, \cdots$, namely at all the open masses.
There is no cancellation from the denominator but
for quite specific values of the external momenta such that $k_1 
\theta k_2$ or $k_3 \theta k_4$ is zero. 

These open string poles deserve some comments. As is well known, 
the form of the Chan-Paton factors of the non-planar amplitude (\ref{calA}) 
of fig.1 indicates that intermediate s-channel states are ``colour''-singlets.
In fig.2 we have drawn a field-theory like diagram representing the 
amplitude $A(1,2,3,4)$. 
\begin{figure}
\label{figura2}
\centerline{\includegraphics[width=6.2cm]{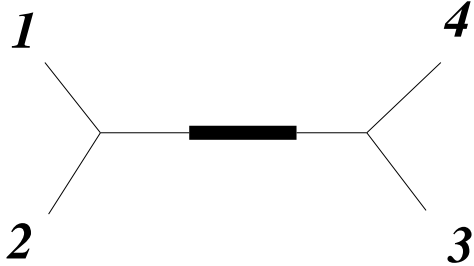}}
\caption{Field-theoretical representation of the amplitude $A(1,2,3,4)$}
\end{figure}
If we single out the first contribution from the closed poles, 
namely the closed-tachyon exchange with residue $\alpha_0$, we can represent it
by a thick line. In addition, double poles arise from integration of the 
$\sin$-functions, which
produce singularities when the variables $\nu_{12}$ and $\nu_3$ approach 
the values zero or one, the corresponding vertex insertions pinching as in fig.2. 
An open string tower travels along the two thin lines, but only singlet states can 
couple to the closed sector. 

Without the $B$-field only a few groups are allowed, according to tree level unitarity: 
$SO(n)$ and $USp(n)$ for non-orientable open strings, $U(n)$ for orientable ones. 
In the non-orientable case, levels $s = -2, 2, \cdots$ (open tachyon, open massive 
graviton, ...) contain singlets, while levels $s = 0, 4, \cdots$ (open vector, ...) 
do not. The pattern of double poles of (\ref{alfa0theta0}) is thereby explained. 

In the orientable case the level $s = 0$ contains a singlet, which is the $U(1)$ 
component in the splitting $U(n)=SU(n)\times U(1)$ and could couple in
principle to the closed channel. Its absence in $a_0$ is explained if one realizes
that the tree level vertex for the production of a $U(1)$ photon from two  
tachyons vanishes for symmetry reasons (see subsection 2.1).

When the $B$-field is present, we find a different situation: first of all, 
the allowed gauge group undergoes a restriction \cite{ad}, and we should consider 
only $U(n)$. This is in agreement with the pattern of poles of (\ref{alfa0totperm}), 
where the double open pole at $s = 0$ is now present. As a matter of fact 
the $B$-field does not alter the nature of the open string spectrum either: Since 
in the case of a gauge group $SO(n)$ or $USp(n)$ the mass-level $s = 0$ does not
contain a singlet component, this would be inconsistent with our findings. 
But groups $SO(n)$ and $USp(n)$ are indeed forbidden. 

For $U(n)$, the $U(1)$ part of the vector still contributes in the closed channel 
and we find the double open string pole at $s = 0$ in (\ref{alfa0totperm}). 
This is due to a ``non-decoupling'' of the $U(1)$ part of $U(n)$, already 
encountered in the presence of the $B$-field for example in the vector 
amplitudes  \cite{bcr}. The simple reason for this phenomenon is that the 
above mentioned tree-level vertex for the production of a $U(1)$ photon from 
two tachyons is now proportional to $\tr [\lambda_1 \lambda_2](p_1 - p_2 )_{\mu} 
\exp [- i p_1 \theta p_2 ]$, which does not vanish after symmetrization
(see again subsection 2.1). The $U(1)$ photon can therefore be produced as a singlet 
and couples to the closed channel. 

In analogy with the factorization found in subsection 2.1, we can now consider
the expression (\ref{alfa0totperm}) and single out the residue at the double pole 
$s = 0$. Exploiting the properties of the Gamma functions, we get:
\beeq
\label{respolodoppio}
&&\alpha_0 (1,2) + \alpha_0 (2,1)\simeq 
\tr [\lambda_1 \lambda_2 ] \, \tr [\lambda_3 \lambda_4 ]{{16}\over{\pi^4}} \,  
{{{\cal{N}}}\over {s^2}}{{1}\over {(-2 - s_{CL} / 4)}} \nonumber \\  
&& \times (k_1 \theta k_2)(k_3 \theta k_4) 
\sin [{{k_1 \theta k_2}\over {2}}] \, 
\sin [{{k_3 \theta k_4}\over {2}}] \, 
\eeeq
where we have for clarity re-inserted the closed tachyon pole and the Chan-Paton 
factors. This part of the amplitude is completely factorisable in the following form:
\beq
\label{factpropag}
{\cal{N}}_1 \, \tr [\lambda_1 \lambda_2 ] \, \sin [{{k_1 \theta k_2}\over {2}}] \, 
{(k_1 - k_2)}_{i} \, {{G^{ij}}\over{s}} \, D_{jl}(K) \, 
{{G^{lm}}\over{s}} {(k_3 - k_4)}_{m} \, 
\sin [{{k_3 \theta k_4}\over {2}}] \, \tr [\lambda_3 \lambda_4 ] \, {\cal{N}}_1, 
\eeq
where 
\beq
\label{propchu}
D_{ij}(K) \, = \, {\cal{N}}_2 \, \theta_{il} K^{l} \, 
{{1}\over {(-2 - s_{CL} / 4)}} \, \theta_{jm} K^{m};   
\eeq
antisymmetry of $\theta$ and total momentum conservation ($K=k_1+k_2=-k_3-k_4$)
have been taken into account.

The factorization above can be easily traced back from fig.2.
In order to complete the proof we should convince the reader that the ``effective 
closed photon
propagator'' has indeed the expression of eq.(\ref{propchu}).
The one-loop non-planar off-shell vectorial 2-point amplitude on the brane 
with momentum $K$
takes the form \cite{bcr}
\begin{eqnarray}\label{2amp} 
  A^{ij} & = & {\cal N}_1 \int_0^\infty [d\tau ] \int_{0}^{1} d\nu
               \, e^{-\frac{\alp\pi}{2\tau }Kg^{-1}K}\nonumber\\
	   & \times & \left[\frac{e^{-\frac{\pi}{4\tau }} 
	      \vartheta_4(\nu,\frac{i}{\tau })}
              {(\frac{1}{\tau })[\eta(\frac{i}{\tau })]^3} 
	      \right]^{-2\alp KG^{-1}K} \times (\rho ' J^{ij}),
\end{eqnarray}
where
\begin{equation}
 [d\tau ] = d\tau (\tau )^{1-d/2} \frac{1}{x} \prod_{n=1}^\infty(1-x^n)^{2-d},
\end{equation}
\begin{equation}
 x = e^{-2\pi \tau } 
\end{equation}
\begin{equation}
 \rho ' = -e^{-2\pi \tau \nu} 
\end{equation}
\begin{equation}
 J = J_0 + J_1 + J_2
\end{equation}
with
\begin{eqnarray}
 J_0 &=& -2\alp \rho {'}  {(\frac{\der I_0}{\der\rho '}|_{\rho=1})}^2
   [G^{ij} K^2- K^i K^j] \nonumber\\
 J_1&=&\frac{i}{\log x} \left(\tilde{K}^i K^j
        \frac{\der I_0}{\der\rho '}+ K^i \tilde{K}^j
	\frac{1}{\rho'}\frac{\der I_0}{\der\rho}\right)|_{\rho=1}\nonumber\\
 J_2&=&-\frac{1}{2\alp(\log x)^2}\frac{1}{\rho '} \tilde{K}^i \tilde{K}^j,
\end{eqnarray}
and
\begin{equation}
  I_0 = \frac{\log^2|\rho\rho {'}|}{2\log x}+\log\left(
  \sqrt{\frac{|\rho|}{|\rho {'}|}}+\sqrt{\frac{|\rho {'}|}{|\rho|}}\right)
  + \log\prod_{n=1}^{\infty}\frac{(1+x^n|\rho/\rho {'}|)(1+x^n|\rho {'}/\rho|)}
  {(1-x^n)^2}.
\end{equation}
We again define
\begin{eqnarray}
  s &=& -K^2 = -K G^{-1} K \\
  s_{CL} &=& -K g^{-1} K
\end{eqnarray}
and set $\alpha'= 1/2$ and $d = 26$.

We look at the singularity at $s=0$. It is easy to realize that only the term 
$J_2$ survives, since $K^2=0$ and $K\cdot (k_1-k_2)=K\cdot (k_3-k_4)=0$.
The effective term in the amplitude is therefrom
\begin{eqnarray}\label{effec}
  A_{eff} &=& {\cal N}_2 \int_0^1 d\nu \int_0^\infty \frac{d\tau}{\tau^2} \, 
\tau^{-12} e^{\frac{\pi}{4\tau }s_{CL}} \nonumber\\  
    & &\times \, \left[ e^{-\pi \tau/12}\prod_{m=1}^\infty(1-e^{-2\pi\tau m})\right]
^{-24}\tilde{K}^i \tilde{K}^j \nonumber\\
    &=& {\cal N}_2 \int_0^\infty \frac{d\tau}{\tau^2} \, \tau^{-12} 
e^{\frac{\pi}{4\tau }s_{CL}} \left[ \eta(i\tau)\right]^{-24}
\tilde{K}^i \tilde{K}^j \nonumber\\
    &=& {\cal N}_2 \int_0^\infty \frac{d\tau}{\tau^2} \,  
e^{\frac{\pi}{4\tau }s_{CL}} \left[ \eta(\frac{i}{\tau})\right]^
{-24}\tilde{K}^i \tilde{K}^j \nonumber\\    
    &=& {\cal N}_2 \sum_{n=0}^{\infty} b_n\int_0^\infty dT\, 
e^{-T\pi(-\frac{s_{CL}}{4}-2+2n)}\tilde{K}^i \tilde{K}^j\nonumber\\
&=& \sum^\infty_{n=0} a_n \frac{1}{-\frac{s_{CL}}{4}-2+2n}\tilde{K}^i \tilde{K}^j,
\end{eqnarray}
where we have used the relation  $\eta(l)=(-il)^{-1/2}\eta(-1/l)$ for the
Dedekind eta-function and expressed the infinite product  
$\prod_{m=1}^\infty(1-e^{\frac{-2\pi m}{\tau}})^{-24}$ as a series 
$\sum_{n=0}^{\infty} b_n e^{-2\pi nT},\, T=\frac1{\tau}$.

The lowest level ($n=0$) exactly coincides with the expression (\ref{propchu}).

\section{Generalization to superstrings}

\subsection{The superstring amplitude}

When considering the corresponding superstring case few novelties will emerge.
The action (\ref{ac}) will be replaced by its supersymmetric extension
\beq \label{supac}
S = S_{bos}+S_{ferm},
\eeq
with
\beq\label{sufer}
S_{ferm}={{1}\over{4\pi {\alpha}'}}\int_{\Sigma}d\sigma d\tau \, (g_{\mu\nu} \, 
{\bar{\xi}}^{\mu} \partial_a \gamma^a \xi^{\nu} - 2 i \pi {\alpha}' B_{ij} \, 
{\epsilon}^{ab} {\bar{\xi}}^i  \partial_a \gamma_b \xi^j ).
\eeq
Here $\xi$ is a two-dimensional Majorana spinor and $\gamma_a$ are the usual
two-dimensional gamma matrices.

We consider the amplitude for the non-planar scattering of four on-shell open 
string vectors, attached on two parallel $Dp$-branes, for instance $D3$-branes, 
in type II superstring 
theory \cite{lm}. Here $p<9$ and one has type IIB for odd $p$, type IIA 
for even $p$. The diagram can be represented as in figure 3 
\begin{figure}
\label{figura3}
\centerline{\includegraphics[width=7cm]{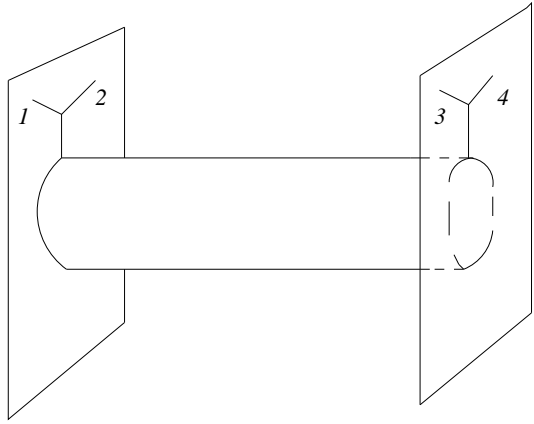}}
\caption{Diagram representing the amplitude $A_S (1,2,3,4)$}
\end{figure}
and the amplitude can be written as 
\beq
\label{calAs}
{\cal{A_S}} = A_S (1,2) \tr [\lambda_1 \lambda_2 ] 
\tr [\lambda_3 \lambda_4 ]+ \mbox{non trivial permutations} 
\eeq
where 
\beeq
\label{As1234}
&&A_S (1,2) = {\cal{N}_{\cal S} } {\cal{K}} \int_0^1 {{dq}\over{q}} q^{{{1}\over {4}}K 
g^{- 1} K} \, {\log [q]}^{(d/2) - 5} \, \exp \left( \vec{Y}^2 / 
\log [q]\right) \nonumber \\
&&\int_0^1 d\nu_1 \int_0^{\nu_1} d\nu_2 \int_0^1 d\nu_3 e^{- {{i}\over {2}} 
k_1 \theta k_2 (1 - 2 \nu_{12} )} e^{{{i}\over {2}} k_3 \theta k_4 (1 - 2 
\nu_{3} )}\nonumber\\
&&{\left[\sin \pi \nu_{12} \prod_1^{\infty} \left( 1 - 2q^{2n}\cos 2\pi 
\nu_{12} + q^{4n}\right)\right]}^{- (s/2)} {\left[\sin \pi \nu_{3} 
\prod_1^{\infty} \left( 1 - 2q^{2n}\cos 2\pi \nu_{3} + q^{4n}\right)\right]}
^{- (s/2)}\nonumber\\
&&{\left[\prod_1^{\infty} \left( 1 - 2q^{2n-1}\cos 2\pi \nu_{13} + 
q^{4n-2}\right)\right]}^{- (u/2)} {\left[\prod_1^{\infty} 
\left( 1 - 2q^{2n-1}\cos 2\pi \nu_{1} + q^{4n-2}\right)\right]}
^{- (t/2)}\nonumber\\   
&&{\left[\prod_1^{\infty} \left( 1 - 2q^{2n-1}\cos 2\pi \nu_{23} + 
q^{4n-2}\right)\right]}^{- (t/2)} {\left[\prod_1^{\infty} 
\left( 1 - 2q^{2n-1}\cos 2\pi \nu_{2} + q^{4n-2}\right)\right]}^{- (u/2)}.
\eeeq
We follow the same notations of the previous sections. Moreover, $d=p+1$ 
is the dimension of the worldvolume of the branes, and ${\vec{Y}}$ 
is their spacelike separation as in \cite{lm,tasi}. The external momenta all lie 
on the branes, which are infinitely heavy objects in perturbation theory, 
breaking translational invariance in the transverse space. We notice the absence 
of the function $f(q^2 )$ when comparing the integrand with the corresponding one in
the bosonic case, due to a cancellation between bosonic and fermionic sectors.
 
The new factor ${\log [q]}^{(d/2) - 5}$ appears instead, which is absent in the 
critical dimension $d=10$; a novelty is the term which accounts for the brane 
separation, which again would be absent for space-filling $D9$-branes. 

However $D9$-branes are not possible here, for two reasons: The first one has to do
with the fact that a $D9$-brane theory is the same as a type I string theory, which 
is endowed with the  $SO(32)$ gauge group. We have already remarked  that such
a group is incompatible with the presence of the $B$-field \cite{bonmardorcrist}. 
The second one is that a $D9$-brane (type I open string) has no NS-NS $B$-field 
in its closed sector, it has a R-R two form which does not couple to the
fundamental superstring according to the simple expression (\ref{supac}). 

A $Dp$-brane on an orientifold of type II has instead $U(1)$ as natural
gauge group (or $U(n)$, if we have a stack of $n$ overlapping branes).
Such a group is allowed in the presence of the $B$-field: We have indeed 
a $B$-field from the closed sector of an orientifold of type II. 

For the sake of generality we consider $n_1$ coincident branes, separated by 
a distance $|{\vec{Y}}|$ from a stack of other $n_2$ coincident branes of 
the same dimension $p+1$. Let the gauge algebra of $\lambda_1,\lambda_2$ 
be $u(n_1 )$ and the one of $\lambda_3,\lambda_4$ $u(n_2 )$. 

${\cal{K}}$ represents the tensorial structure of the amplitude: it 
originates from the trace over the fermionic part of the gauge boson 
vertex operator, saturated with the relevant polarization vectors \cite{lm}. 
One finds
\beq
\label{K}
{\cal{K}} = {M_{1}}_{i}^{j} {M_{2}}_{j}^{l} {M_{3}}_
{l}^{m} {M_{4}}_{m}^{n} - {{1}\over {4}} {M_{1}}_{ij} 
{M_{2}}^{ji} {M_{3}}_{lm} {M_{4}}^{ml} + 2 \,\mbox{permutations} ,
\eeq
where ${M_{r}}^{ij} = {e_{r}}^{i} {k_{r}}^j - {e_{r}}^j
{k_{r}}^i$, ${e_r}$ being the polarization vectors; indices are raised 
and lowered with the open string metric. 

The integrand has almost the same form of the bosonic case, apart from the 
different $q$-structure and the already mentioned changes.
We can again write
\beq
\label{sommapolis2theta}
A_S (1,2) = \sum_{n=0}^{\infty} f_n (s_{CL} , d, |{\vec{Y}}|) \, 
{\alpha^S_n}(s,t,u, k_1 \theta k_2 , k_3 \theta k_4 ),
\eeq
with
\beq
\label{KG}
f_n (s_{CL} , d, |{\vec{Y}}|) = \int_0^1 \, dq \, q^{- 1 + n - (s_{CL} / 4)} \, 
\left({\log [q]}\right)^{(d/2) - 5} \, \exp \left( {\vec{Y}}^2 / \log [q]\right).  
\eeq
The expression of $\alpha^S_1$ is essentially the same as the bosonic one 
eq.(\ref{alfa1finale}), with $\cal{N_S}$ replacing $\cal{N}$, the extra 
tensorial factor 
${\cal{K}}$ and the shifts $(s,t,u)\rightarrow (s-4,t-4,u-4)$. 
In particular, it is again odd under the exchange of $k_1\leftrightarrow k_2$ and 
therefrom vanishing when suitably symmetrized. According to Appendix B
only even terms survive
\beq
\label{sommapolis2theta2}
A_S (1,2) = \sum_{n=0}^{\infty} f_{2n} (s_{CL} , d, |{\vec{Y}}|) \, 
{\alpha^S_{2n}}(s,t,u, k_1 \theta k_2 , k_3 \theta k_4 ).
\eeq
Let us give a closer look at the coefficients $f_{2n}$. We see that
\beq
\label{KG2}
f_{2n} (s_{CL} , d, |{\vec{Y}}|) \, = \, - 4 \, i \, {(- {{1}\over {\pi }})}
^{{{d-10}\over {2}}} \, G_{9 - p} ({\mu}_{2n}^2 \, ; \, |{\vec{Y}}|),
\eeq
where $G_{9 - p} ({\mu}_{2n}^2 \, ; \, |{\vec{Y}}|)$ is nothing but
the propagator kernel for a particle of mass ${\mu}_{2n}^2 = 8n - s_{CL}$ 
at a spatial distance $|{\vec{Y}}|$
\beq
\label{KleinGordon}
G_{9-p} (\mu_{2n}^2 \, ; \, |{\vec{Y}}|) \, = \, {{i}\over {{(2 \pi )}^{9-p} }} 
\int d^{9-p} \vec{k} \, \, {{\exp [i \vec{k}\cdot \vec{Y} ]}\over {{\vec{k}}^2 
+ {\mu}_{2n}^2 - i \epsilon }}.
\eeq
If we recall that $s_{CL}$ gets contribution from the momenta lying on the branes, 
and we split the momentum of a generic bulk particle as $ K_{tot}^2 = 
{\vec{k}}^2 - s_{CL}$, the above coordinate propagator can be thought as 
representing the Fourier transform of a momentum space propagator 
$\frac 1 { K_{tot}^2 + M^2-i\epsilon}$ 
for a spatial distance $|\vec{Y}|$ (the position of the two 
fixed branes). Then consistency requires $M^2 = 8n$, which is precisely
the closed string spectrum. We stress once again that we need to suitably 
symmetrize the amplitude in order to recover the correct pattern of closed 
string poles.   

The factorization in eq.(\ref{sommapolis2theta}) deserves a comment. The residues on
the closed string poles possess a tensorial structure on their own, which depends
on the level $n$ we consider. When coupled to the branes, only the vector components 
along the branes survive and do not undergo the integration in (\ref{KleinGordon}).
Their contribution is included in $\alpha^S_n$.

Now we turn our attention to the contribution related to the massless closed pole, 
$\alpha^S_0$: After symmetrization we obtain the same result as in 
eq.(\ref{alfa0totperm}), with $s \rightarrow s - 4$ and the factors 
$\cal{N_S}$ and $\cal{K}$
\beq
\label{alfa0stotperm}
\alpha^S_0 (1,2) + \alpha^S_0 (2,1)= {{{\cal{N_S}} {\cal{K}} \, \, 2^{s} \, 
\, \Gamma^2 (1 - {{s}\over {2}})}\over {\Gamma (1 - {{s}\over {4}} + {{k_3 
\theta k_4 }\over {2 \pi}}) \Gamma (1 - {{s}\over {4}} - {{k_3 \theta k_4 }
\over {2 \pi}}) \Gamma (1 - {{s}\over {4}} + {{k_1 \theta k_2 }\over {2 \pi}}) 
\Gamma (1 - {{s}\over {4}} - {{k_1 \theta k_2 }\over {2 \pi}})}}.
\eeq

When $\theta$ vanishes, this expression becomes
\beq
\label{astotperm}
a^S_0 (1,2) + a^S_0 (2,1)= {{{\cal{N_S}} {\cal{K}} \, 
\Gamma^2 ({{2-s}\over {4}})}\over {\pi\Gamma^2 (1 - {{s}\over {4}})}}. 
\eeq
There are poles at $s = 2, 6, \cdots$, again in the correct traditional pattern.

When $B$ is turned on, all the poles in the numerator of eq.(\ref{alfa0stotperm}),
namely at $s = 2, 4, 6, 8,..$ are present, the ones at $s = 4, 8 ,...$ being 
no longer canceled by the denominator, apart from exceptional configurations 
of the momenta.

We notice a difference with the bosonic case: the open string pole at $s = 0$ is 
always absent. Without the $B$-field, this massless pole is  forbidden by 
supersymmetry (the vanishing of the one loop three-vectors amplitude due to a 
non-renormalization of the three-vector vertex). The presence of $B$ does preserve 
this constraint: In particular, any vector amplitude on the annulus with less 
than eight spinors, such as the three point amplitude, continues to be zero \cite{lm}. 

\subsection{String singularities and unitarity in the Seiberg-Witten limit}

We restore in the amplitude (\ref{As1234}) the dependence on $\alpha'$ and introduce
the notation 
$K\circ K\equiv -K\th G\th K$. The quantity $\frac{|\vec{Y}|}{2\pi \alpha'}\equiv m$
behaves like a mass scale in the theory; it is interpreted as the mass of the ground
states of open strings stretching between the branes \cite{lm,tasi}. We choose
$\th $ block-diagonal and distinguish between the ``magnetic''
case, where $K\circ K$ is positive definite ($\th $ has only spatial components), 
and the ``electric'' case where, for simplicity, $\th $ is chosen with the
only non-vanishing component $\th_E=\th_{01}$. 

We are now interested in exploring the analytic behaviour of the amplitude 
(\ref{A1234}) as a function of the variable $s$ at suitable fixed values
of the other kinematic variables.
This behaviour will be
eventually compared with the one of the corresponding amplitude in the
noncommutative field theory obtained by performing the Seiberg-Witten limit.

To this purpose we express the variable $s_{CL}$ in term of $s$
\beq \label{essecl}
s_{CL} = s-\frac1{4\pi^2 \alpha'^2} K\circ K.
\eeq

The amplitude has a rich pattern of singularities, due to the fact that it describes
simultaneously several physical processes in different kinematic regions.
Besides the double poles at $s = 2n,\,\, n=1,2,...$ we have already mentioned,
it exhibits a branch point at $s = 4m^2$ with a cut along the positive
real axis. It is the unitarity cut related to the open strings attached to
the branes. In the representation
\beeq \label{ampl24tach2R}
  A_{S} & = & \mathcal{N}_{\cal S}\mathcal{K}
              \int_0^\infty \frac{dl}{l} l^{4-d/2}  
	      e^{-\frac{\pi^2\alp^2}{l} K g^{-1} K} e^{-\frac{\vec{Y}^2l}{(2\pi\alp)^2}}
	      e^{-\frac{i}{2}k_1\th k_2}
	      e^{\frac{i}{2}k_3\th k_4}\int_0^1 d\nu_1 \int_0^{\nu_1} d\nu_2 \nonumber \\
	  & \times & \int_0^1 d\nu_3
	      e^{i(\nu_{12})k_1\th k_2}
	      e^{-i\nu_3k_3\th k_4}
	      \prod_{n=1}^{2}
	      \prod_{m=3}^{4}\left[\frac{e^{-\frac{\pi^2\alp}{2l}} 
	      \vartheta_4(\nu_{nm},\frac{2\pi i\alp}{l})}
              {(\frac{2\pi\alp}{l})
	      [\eta(\frac{2\pi i\alp}{l})]^3} \right]^{2\alp
		k_nG^{-1} k_m}  \nonumber \\
          & \times  & 
	      \left[\frac{\vartheta_1
	      (\nu_{12},\frac{2\pi i\alp}{l})}
	      {(\frac{2\pi\alp}{l})
	      [\eta(\frac{2\pi i\alp}{l})]^3} \right]^{2\alp
	      k_1G^{-1} k_2}  \left[\frac{\vartheta_1
	      (\nu_3,\frac{2\pi i\alp}{l})}
	      {(\frac{2\pi\alp}{l})
	      [\eta(\frac{2\pi i\alp}{l})]^3} \right]^{2\alp
	      k_3G^{-1} k_4},  
\eeeq
it originates from the integration over high 
values of $l$ as explained in Appendix C.  

Low values of $l$ describe instead the effect of the closed string exchange
between the branes. The ensuing singularities are particularly interesting
in the Seiberg-Witten limit.
Eq.(\ref{KG}) leads to the following representation
\beq
\label{KGE}
f_n (s_{CL} , d, |{\vec{Y}}|) \propto \Big(\frac{2\pi m \alpha'}{n-
\frac{\alpha' s_{CL}}4} \Big)^{d/2-4} K_{d/2-4}\Big(4\pi \alpha' m
\sqrt{n-\frac{\alpha' s_{CL}}4}\Big), \qquad n=0,1,...\qquad.
\eeq 

A branch point occurs when the argument of the square root becomes negative.
If we trade the variable $s_{CL}$  for $s$ we obtain the conditions
\beq \label{branch}
s \ge \frac{4n}{\alpha'}+\frac1{4\pi^2 \alpha'^2} K\circ K, \quad n=0,1,... \, .
\eeq
If $\th$ is magnetic, $K\circ K$ is positive definite. We find thereby cuts 
along the positive real axis.

If instead $\th$ is electric,
\begin{equation}
  K\circ K = \th_E^2(K_0^2-K_1^2) = \th_E^2 s + \th_E^2 K_T^2 ,
\end{equation}
where $K_T^2 \equiv K_2^2 + K_3^2 + ... + K_{d-1}^2$.
Eq.(\ref{branch}) then becomes
\begin{equation}\label{elepo}
   s_n(1-\frac{\th_E^2}{4\pi^2\alp^2})=\frac{\th_E^2}
  {4\pi^2\alp^2} K_T^2 + \frac{4n}{\alp},
\end{equation}
that, if we define $\tilde{E} = \frac{E}{E_{cr}} =  \frac{\th_E}{2\pi\alp}$, 
can be rewritten as
\begin{equation}\label{elecr}
  s_n(1-\tilde{E}^2) = \tilde{E}^2 K_T^2 +
  \frac{4n}{\alp}.  
\end{equation}
Thereby the string theory is stable whenever
$|\tilde{E}|<1$, namely
\begin{equation}\label{flip}
s>\frac{\tilde{E}^2}{1-\tilde{E}^2} K_T^2 +\frac{1}{1-\tilde{E}^2} 
\frac{4n}{\alp} >0.
\end{equation}

When $E$ overcomes $E_{cr}$, the theory exhibits a perturbative instability 
of a tachyonic type; the vacuum is likely to decay into a suitable 
configuration of branes. 

\smallskip

We turn now our attention to the behaviour of such an amplitude in the Seiberg-
Witten limit.
It is well known that open strings in presence of an antisymmetric constant background
are effectively described at low energy by  noncommutative field theories. 
If in the amplitude (\ref{As1234}) we perform the limit 
$\alp\rightarrow 0, g_{ij}\simeq\alp^2$, keeping  
$G, \th$ and $m=\frac{|\vec{Y}|}{2\pi \alp}$ fixed, and suitably rescaling
the string coupling constant (see \cite{ad,sw}), we get \cite{lm}
\beeq
\label{Asalp}
&&A = {\cal{K}} \, N \,\delta^{d}\lp \sum_{m=1}^4 k_m \rp
\int_0^\infty \frac{dl}{l} l^{4-\frac{d}{2}}  e^{-\frac{1}{4l}K\circ K -m^2l} 
e^{- {{i}\over {2}}k_1 \theta k_2 } e^{{{i}\over {2}} k_3 \theta k_4 }\nonumber \\
&&\int_0^1 d\nu_1 \int_0^{\nu_1} d\nu_2 \int_0^1 d\nu_3 e^{- i k_1 
\theta k_2 \nu_{12}} e^{i k_3 \theta k_4 \nu_{3}} \prod_{i<j=1}^4  
e^{l\nu_{ij}(1-\nu_{ij})k_iG^{-1}k_j} .
\eeeq
This expression corresponds to the sum of three non-planar diagrams.

One can easily check that in such a limit the open string (double) poles
decouple. They are indeed absent in eq.(\ref{Asalp}). The right-hand cut for
$s\ge 4m^2$, which is
present (with a different discontinuity )
also in commutative field theories due to intermediate on shell states,
obviously survives.

A different fate occurs to the branch points related to the closed string singularities.
In the magnetic case, if we look at the eq.(\ref{branch}), we see that in the limit
$\alp \to 0$, they decouple as well.

If instead $\theta$ is of ``electric'' type,
the Seiberg-Witten limit $\alp\to 0, \th_E$ fixed, 
forces $\tilde E \to \infty >1$. Even before reaching the noncommutative field 
theory limit,
the string is pushed into its region of instability owing to the
flipping of the branch point (see eq.(\ref{elecr})). In the meanwhile the closed
branch points $s_n$ get closer and closer to $-K_T^2$ (see eq.(\ref{elepo})).  
No wonder then that the amplitude in the
resulting noncommutative theory eventually exhibits un unphysical cut
of tachyonic type for $s<-K_T^2$.

\section{Concluding remarks}

The main purpose of this work was to explore up to one loop the perturbative 
spectra and the related factorization properties of (super)string amplitudes in the 
presence of a constant background field $B$. The final goal was to match
peculiar (and sometimes pathological) features of noncommutative field
theories, derived by means of the Seiberg-Witten limit, with the corresponding
characteristics of the parent theory. This analysis was first performed
in a simpler bosonic string context, in spite of the well-known 
difficulties due to the presence of tachyons and to the large number of 
extra dimensions.

Closed and open string poles occurring in the four open-string scattering
amplitude were analyzed. After the required symmetrization, the closed string spectrum
turns out to be unaffected by the presence of $B$, as expected from unitarity.
Extra open string poles are instead present when $B$ is switched on, this
field providing further possible structures and therefrom a partial
symmetry loss. Remarkable factorization properties occurring in the
residues on the poles were found in full agreement with expected patterns.

This analysis was then extended to the more satisfactory superstring context.
Here two stacks of $N$ (lower dimensional) branes at a relative distance $\vec{Y}$
were considered in order to deal with the Chan-Paton group $U(N)$.
This distance (divided by the string slope $\alp$) provides us with a mass scale,
which can be interpreted as an IR or an UV effect,
according to the different interpretations of fig.3  \cite{tasi}.

When the background field is of an ``electric'' type, there is a critical
value of the electric field beyond which the string undergoes a classical
perturbative instability and starts developing tachyonic poles.
In the Seiberg-Witten limit this ``electric'' string is necessarily
pushed into its instability region; the resulting noncommutative field
theory turns out to exhibit an unphysical cut whose presence was noticed since a long
time \cite{gm,bgnv}.

A future step would be to explore this phenomenon in the more ambitious
context of string field theories; this might allow to go beyond 
on-shell amplitudes and perhaps also beyond perturbation theory in the search
of new, more satisfactory ground states.

\acknowledgments{We thank G. De Pol for stimulating discussions. One of us (A.T.)
acknowledges a grant from the ``Deutscher Akademischer Austauschdienst'' 
under the reference 
A0339307 (sec.14) and
wishes to thank Professor Harald Dorn for discussions and for the hospitality he enjoyed at the
Institute of Physics of the Humboldt-University in  Berlin, while part of this work 
was done.}

\appendix
\section{The four-point amplitude in the q-variable}

We report here the relevant formulas which connect the $\tau$-variable representation
(\ref{ampl}) and the $q$-variable representation (\ref{A1234}) 
for the four-point amplitude.

Starting from the (\ref{ampl}), we first perform the rescaling 
$t = 2 \pi \alpha' \tau$ and $\nu_i  = {{y_i}\over{2 \tau}}$. 
As mentioned in the text, we work in the gauge $\nu_4 = 0$. 
After such a rescaling, in (\ref{ampl}) $q$ becomes equal to 
$\exp [- 2 \pi^2 \alpha' / t]$. 

We use the well-known expansions of the Jacobi $\theta$-functions \cite{gsw}:
\beq
\label{2}
\theta_1 [\nu_{ij}, {{2 \pi i \alpha'}\over{t}}] = 2 \, f(q^2 ) \, q^{1/4} \, 
\sin \pi \nu_{ij} \, \prod_{n=1}^{\infty} (1 - 2 q^{2n} \cos 2 \pi \nu_{ij} \, 
+ q^{4n} ) ,
\eeq
\beq
\label{3}
\theta_4 [\nu_{ij}, {{2 \pi i \alpha'}\over{t}}] = f(q^2 ) \, 
\prod_{n=1}^{\infty} (1 - 2 q^{2n-1} \cos 2 \pi \nu_{ij} \, + q^{4n-2} ) ,
\eeq
where, as in the main text $\nu_{ij} = \nu_i - \nu_j$.

With the definition $\eta (s) = x^{{{1}\over{24}}} 
\prod_{m=1}^{\infty} (1 - x^m )$ with $x = \exp [2\pi i s]$, we can write 
\beq
\label{4}
D[{{t}\over{2 \pi \alpha'}}] = {{- \log q}\over{\pi}} \, q^{1/4} \, f^3 (q^2 ) ,
\eeq
where $f(q^2) = \prod_{n=1}^{\infty} (1 - q^{2n} )$. 

The duality transformation $\eta (s) = {(- i s)}^{-1/2} \eta (- 1/s)$ leads to
\beq
\label{5}
\eta ({{i t}\over {2 \pi \alpha'}}) = {({{-\pi}\over{\log q}})}^{-1/2} \, 
q^{1/12} \, f(q^2 ).
\eeq
Finally, changing variable from $t$ to $q$ and going on-shell in 
$d=26$ (which in particular implies $\sum_{i<j} k_i k_j = - 4$ 
if $\alpha' = 1/2$ and $G$ is the Minkowski metric as in our conventions), 
we obtain eq.(\ref{A1234}). 

We stress that, like in the case without the B-field, only in the critical 
dimension and going 
``on-shell'', the logarithms occurring in the measure exactly cancel against 
the ones coming from the propagator insertions, so that eventually only closed 
string poles ensue, in compliance with unitarity as explained in the text.

\section{Cancellation of poles for odd values of n}

The $n$th-order coefficient of the sum (\ref{sommapoli2theta}) is 
\begin{eqnarray}\label{alpR}
  && \sum_{h}^n \sum_{\a_1+\a_2+\a_3+\a_4=h} 
\int_0^1 d\nu_3 \int_0^1 dp \int_p^1 d\nu_1
	e^{-\frac{i}{2}k_1\th k_2} e^{\frac{i}{2}k_3\th k_4}  \nonumber\\
	&&e^{i k_1\th k_2\, p} e^{-i k_3\th k_4\,\nu_3} 
	A_h(\sin\pi p,\sin\pi\nu_3;s,t,u)  \nonumber\\
	&&f_1(u) [\cos 2\pi(\nu_1-\nu_3)]^{\a_1}\,   
	f_2(t) [\cos 2\pi\nu_1]^{\a_2} \cdot \nonumber\\ 
  	&&f_3(t) [\cos 2\pi(\nu_1-p-\nu_3)]^{\a_3} \, 
	f_4(u) [\cos 2\pi(\nu_1-p)]^{\a_4},  
\end{eqnarray}
where $h$ runs over values with the same parity of $n$ and $\alpha_i$ are
non-negative integer numbers whose dependence in $f_a, a=1,...,4$ is understood;
$A_h$ is a symmetric function of $t$ and $u$.

This sum can be rewritten by symmetrizing over the variables $\a_i$:
\begin{eqnarray}\label{alpsimR}
	&&\frac{1}{8}\sum_{h}^n \sum_{\a_1+\a_2+\a_3+\a_4 = h} 
	\int_0^1 d\nu_3 \int_0^1 dp \int_p^1 d\nu_1e^{-\frac{i}{2}k_1\th k_2} 
	e^{\frac{i}{2}k_3\th k_4} \nonumber\\
	&&e^{i k_1\th k_2\, p} e^{-i k_3\th k_4\,\nu_3} A_h(\sin\pi p,
        \sin\pi\nu_3;s,t,u)\, [f_a(u)f_b(t) {\cal M}_1 + f_a(t)f_b(u) {\cal M}_2],
\end{eqnarray}
where
\beeq \label{trigo}
&&{\cal M}_1= \Big([\cos 2\pi(\nu_1-\nu_3)]^{\a_1} [\cos 2\pi(\nu_1-p)]^{\a_4}
+[\cos 2\pi(\nu_1-\nu_3)]^{\a_4} [\cos 2\pi(\nu_1-p)]^{\a_1}\Big) \cdot \nonumber  \\ 
&&\cdot\Big([\cos 2\pi(\nu_1-p-\nu_3)]^{\a_3}[\cos 2\pi\nu_1]^{\a_2} 
+[\cos 2\pi(\nu_1-p-\nu_3)]^{\a_2}[\cos 2\pi\nu_1]^{\a_3}\Big),
\eeeq
${\cal M}_2$ is obtained by performing the exchange $\alpha_1\leftrightarrow
\alpha_2, \, \alpha_4 \leftrightarrow \alpha_3$ in ${\cal M}_1$.

We want to verify that this expression vanishes when $n$ is odd. It is enough
to prove that  
\beeq\label{azzera}
&&\int_0^1 d\nu_3 \int_0^1 dp \int_p^1 d\nu_1e^{-\frac{i}{2}k_1\th k_2} 
	e^{\frac{i}{2}k_3\th k_4} e^{i k_1\th k_2\, p} 
        e^{-i k_3\th k_4\,\nu_3} \cdot \nonumber \\
&&\cdot A_h(\sin\pi p,
        \sin\pi\nu_3;s,t,u)\, [f_a(u)f_b(t) {\cal M}_1+ f_a(t)f_b(u){\cal M}_2] = 0 
\eeeq	
for a generic choice of 
$\a_1,\a_2,\a_3,\a_4$ provided that $\a_1+\a_2+\a_3+\a_4$ is odd, after the
symmetrization $k_1 \leftrightarrow k_2$.

Thanks to the symmetries $ \a_1 \leftrightarrow \a_4$ and $\a_2
\leftrightarrow \a_3$, we can set $\a_1 < \a_4$ and $\a_2 < \a_3$
without loss of generality.

Let us consider ${\cal M}_1$. It can be written as
\beeq \label{ast11R}
	&&{\cal M}_1= \nonumber \\
        && [\cos 2\pi(\nu_1-\nu_3) \cos 2\pi(\nu_1-p)]^{\a_1} \cdot 
	 [(\cos 2\pi(\nu_1-\nu_3))^{\a_4-\a_1} +
	 (\cos 2\pi(\nu_1-p))^{\a_4 - \a_1} ] \cdot \nonumber \\ 
	&& [\cos 2\pi(\nu_1-p-\nu_3)\cos 2\pi\nu_1]^{\a_2} \cdot 
	 [(\cos 2\pi(\nu_1-p-\nu_3))^{\a_3-\a_2} +
	 (\cos 2\pi\nu_1)^{\a_3-\a_2} ],
\eeeq
which, after using standard algebraic and trigonometric identities and 
dropping the irrelevant factor 
$2^{-\a_1-\a_2}$ common to ${\cal M}_1$ and ${\cal M}_2$, becomes
\beeq \label{ast12R}
	&&{\cal M}_1= [\cos 2\pi(2\nu_1-p-\nu_3) + \cos 2\pi(p-\nu_3)]^{\a_1}
	\cdot 
	[\cos 2\pi(2\nu_1-p-\nu_3)+\cos 2\pi(p+\nu_3)]^{\a_2} \nonumber \\
	&&\cdot \{[\cos \pi(2\nu_1-p-\nu_3-(p-\nu_3))]^{\a_4-\a_1} 
	+ [\cos \pi(2\nu_1-p-\nu_3+(p-\nu_3)) ]^{\a_4 - \a_1}\} 
	\cdot \nonumber \\ 
	&&\cdot \{ [\cos \pi(2\nu_1-p-\nu_3-(p+\nu_3))]^{\a_3-\a_2} 
	+ [\cos \pi(2\nu_1 -p-\nu_3+(p+\nu_3))]^{\a_3-\a_2}\}. 
\eeeq
After some algebra, this quantity
takes the form
\beq \label{split}
{\cal M}_1= {\cal F}_1\cdot {\cal F}_2\cdot {\cal F}_3\cdot {\cal F}_4,
\eeq
with
\beeq\label{f}
{\cal F}_1 &=& \sum_{n_1+m_1=\a_1}a_{n_1m_1}\cos^{n_1} 2\pi(p-\nu_3)
	(1-2\sin^2 \pi(2\nu_1-p-\nu_3))^{m_1}, \nonumber \\
{\cal F}_2 &=& \sum_{n_2+m_2=\a_2}a_{n_2m_2}\cos^{n_2} 2\pi(p+\nu_3)
	(1-2\sin^2 \pi(2\nu_1-p-\nu_3))^{m_2},  \\
{\cal F}_3 &=& [\cos \pi(2\nu_1-p-\nu_3)\cos \pi(p-\nu_3) 
	+ \sin \pi(2\nu_1-p-\nu_3)\sin \pi(p-\nu_3) ]^{\a_4-\a_1} \nonumber \\
&+&[\cos \pi(2\nu_1-p-\nu_3)\cos \pi(p-\nu_3) 
	- \sin \pi(2\nu_1-p-\nu_3)\sin \pi(p-\nu_3) ]^{\a_4-\a_1}, \nonumber \\ 
{\cal F}_4 &=& [\cos \pi(2\nu_1-p-\nu_3)\cos \pi(p+\nu_3) 
	+ \sin \pi(2\nu_1-p-\nu_3)\sin \pi(p+\nu_3) ]^{\a_3-\a_2} 
	\nonumber \\ 
	&+&[\cos \pi(2\nu_1-p-\nu_3)\cos \pi(p+\nu_3) 
	- \sin \pi(2\nu_1-p-\nu_3)\sin \pi(p+\nu_3) ]^{\a_3-\a_2}.\nonumber  
\eeeq	
We consider first the case in which $(\a_4-\a_1)$ is odd and, consequently,
$(\a_3-\a_2)$ is even.
\noindent
${\cal F}_3$ takes the form
\begin{eqnarray}\label{par3R}
 {\cal F}_3 &=&  [I_1+J_1]^{\a_4-\a_1} 
  + [I_1-J_1]^{\a_4-\a_1}  \\
  &=&  \sum_{n_4+m_4 = \a_4-\a_1}    
  b'_{n_4m_4}[I_1^{n_4}J_1^{m_4} +
  I_1^{n_4}(-J_1)^{m_4}]\nonumber\\
  &=&  \sum_{l_4,h_4;\, 2l_4+1+2h_4 = \a_4-\a_1}    
  b_{l_4h_4} [I_1^{2l_4+1}J_1^{2h_4}]. \nonumber
\end{eqnarray}
In the sum only the even powers of $J_1$ survive, and consequently the
odd ones of $I_1$ ($(\a_4-\a_1)$ is odd).
\noindent
${\cal F}_4$  is analogous, with the difference that $(\a_3-\a_2)$
is even
\begin{eqnarray}\label{par4R}
  {\cal F}_4  &=&  [I_2+J_2]^{\a_3-\a_2} 
  + [I_2-J_2]^{\a_3-\a_2}  \\
  &=&  \sum_{l_3,h_3;\, 2l_3+2h_3 = \a_3-\a_2}    
  b_{l_3h_3} [I_2^{2l_3}J_2^{2h_3}]. \nonumber
\end{eqnarray}
As a consequence ${\cal M}_1$ becomes
\begin{eqnarray}\label{ast14R}
  {\cal M}_1 &=&\sum_{n_1m_1,n_2m_2,l_3h_3,l_4h_4}
    a_{n_1m_1}a_{n_2m_2}b_{l_3h_3}b_{l_4h_4} \cdot\nonumber\\
  &&[\cos2\pi(p-\nu_3)]^{n_1} \cdot [\cos2\pi(p+\nu_3)]^{n_2} 
    \cdot [\cos\pi(p-\nu_3)]^{2l_4+1}\cdot\nonumber\\
  &&[\cos\pi(p+\nu_3)]^{2l_3}\cdot[\sin\pi(p-\nu_3)]^{2h_4}\cdot
    [\sin\pi(p+\nu_3)]^{2h_3}\cdot \\
  &&[1-2\sin^2\pi(2\nu_1-p-\nu_3)]^{m_1+m_2}
    \cdot[1-\sin^2\pi(2\nu_1-p-\nu_3)]^{l_3+l_4}\cdot \nonumber\\
  &&[\sin\pi(2\nu_1-p-\nu_3)]^{2(h_3+h_4)}\cdot \,\cos\pi(2\nu_1-p-\nu_3)\cdot 
     \nonumber
\end{eqnarray}
The integral in $\int_p^1d\nu_1$ of (\ref{ast14R}) is the sum of
several integrals of the type:
\begin{eqnarray}
  &&g(\nu_3,p)\int_p^1 [\sin\pi(2\nu_1-p-\nu_3)]^{2\gamma}
  cos\pi(2\nu_1-p-\nu_3)d\nu_1\nonumber\\
  &&= \,-\frac{g(\nu_3,p)}{2\pi(2\gamma+1)}[\sin^{2\gamma+1}\pi(p+\nu_3) 
	+ \sin^{2\gamma+1}\pi(p-\nu_3)],
\end{eqnarray}
with non-negative $\gamma$.

In turn the integral over $\nu_1$ of (\ref{ast14R}) can be expressed as the sum of
terms like the following ones with different coefficients:
\begin{eqnarray}\label{addst1R}
{\cal T}_1=  &&[\cos2\pi(p-\nu_3)]^{n_1} \cdot [\cos2\pi(p+\nu_3)]^{n_2} 
  \cdot [\cos\pi(p-\nu_3)]^{2l_4+1}\cdot \nonumber\\
  &&[\cos\pi(p+\nu_3)]^{2l_3}
   \cdot[\sin\pi(p-\nu_3)]^{2h_4}\cdot
   [\sin\pi(p+\nu_3)]^{2h_3}\cdot \nonumber\\
  &&[\sin^{2\gamma+1}\pi(p+\nu_3) + 
  \sin^{2\gamma+1} \pi (p-\nu_3)] \nonumber\\
  && \nonumber \\ 
  &&=\,[\cos2\pi p\cos2\pi\nu_3 +\sin2\pi p\sin2\pi\nu_3 ]^{n_1} \cdot
  [\cos2\pi p\cos2\pi\nu_3 -\sin2\pi p\sin2\pi\nu_3 ]^{n_2} \cdot\nonumber\\
  &&[\cos\pi p\cos\pi\nu_3 +\sin\pi p\sin\pi\nu_3 ]^{2l_4+1} \cdot
  [\cos\pi p\cos\pi\nu_3 -\sin\pi p\sin\pi\nu_3 ]^{2l_3}\cdot \nonumber\\
  &&[\sin\pi p\cos\pi\nu_3 -\cos\pi p\sin\pi\nu_3 ]^{2h_4} \cdot
  [\sin\pi p\cos\pi\nu_3 +\cos\pi p\sin\pi\nu_3 ]^{2h_3} \cdot\nonumber\\
  &&\{[\sin\pi p\cos\pi\nu_3 +\cos\pi p\sin\pi\nu_3 ]^{2\gamma+1} +
  [\sin\pi p\cos\pi\nu_3 -\cos\pi p\sin\pi\nu_3 ]^{2\gamma+1}\} \cdot\nonumber\\
    &&\\
  &&\equiv [A+B]^{n_1}  \cdot [A-B]^{n_2} \cdot [C+D]^{2l_4+1} 
  \cdot [C-D]^{2l_3}\cdot\nonumber\\
  &&\,\,\,\,\,[E-F]^{2h_4} \cdot [E+F]^{2h_3} \cdot \{G\}. 
  \nonumber
\end{eqnarray}
We remember that ${\cal M}_1$ is multiplied by $[f_a(u) \cdot f_b(t)]$.
\noindent
To each of the terms like ${\cal T}_1$ an analogous one
corresponds,  belonging to the
integral in $\nu_1$ of ${\cal M}_2$, with 
the same coefficient, given by
\begin{eqnarray}\label{addst2R}
{\cal T}_2=  &&[A+B]^{n_2}  \cdot [A-B]^{n_1} \cdot [C+D]^{2l_3} \cdot [C-D]^{2l_4+1}
\cdot\nonumber\\
  &&\,\,\,\,\,[E-F]^{2h_3} \cdot [E+F]^{2h_4} \cdot \{G\} ,
\end{eqnarray}
and multiplied by $[f_a(t) \cdot f_b(u)]$.

\smallskip

Now we perform the integral 
\begin{eqnarray}\label{intaddstR1}
  &&\int_0^1d\nu_3 e^{\frac{i}{2}k_3\th k_4} e^{-i k_3\th
	k_4\,\nu_3} \int_0^1dp e^{-\frac{i}{2}k_1\th k_2}
	e^{i k_1\th k_2\, p} \nonumber\\
	&&\,\,\,\,\,\,\,\,\,\,\,A_h(\sin\pi p,\sin\pi\nu_3;s,t,u) 
	[f_a(u) \cdot f_b(t){\cal T}_1 + f_a(t) \cdot f_b(u) {\cal T}_2].
\end{eqnarray}
These integrals are of the type 
\begin{eqnarray}\label{a1}
   \int_0^1 dx e^{-\frac{i}{2}a}e^{i a x}  {(\sin \pi x )}^b & = & 
   {{2^{-b} \Gamma (1 + b) }\over {\Gamma (1 + {{b}\over {2}} - 
{{a}\over {2 \pi }}) \Gamma (1 + {{b}\over {2}} + {{a}\over {2 \pi }})}}\nonumber\\
&&\\
\int_0^1 dx e^{-\frac{i}{2}a}e^{i a x}  {(\sin \pi x )}^b  
\cos \pi x & = & {{-i a 2^{-b-1} \Gamma (1 + b) }\over {\pi 
\Gamma ({{3}\over {2}} + {{b}\over {2}} - {{a}\over {2 \pi }}) 
\Gamma ({{3}\over {2}} + {{b}\over {2}} + {{a}\over {2 \pi }})}}.\nonumber
\end{eqnarray}
We concentrate our attention on the integrals in $p$, because they are even
functions of $k_1\th k_2$ when the integrand contains even powers of
$\cos\pi p$ and are odd otherwise. 
The integral in $\nu_3$ does involve $k_1$ and $k_2$ only in their symmetric
combination $s$.  

\smallskip

We note that the factor $G$ 
appears in both (\ref{addst1R}) and (\ref{addst2R}), and that it is
a function of even powers of $\cos\pi p$; we could prove it by repeating the
arguments we used in (\ref{par3R}). Therefore we
focus on the quantity
\begin{eqnarray}\label{relevexpR1}
{\cal Q}=  &&[A+B]^{n_1} [A-B]^{n_2} [C+D]^{2l_4+1} [C-D]^{2l_3}[E-F]^{2h_4}
\cdot\nonumber\\ 
  &&[E+F]^{2h_3} \cdot [f_a(u) \cdot f_b(t)] +\nonumber\\
  &&+ \,[A+B]^{n_2} [A-B]^{n_1} [C+D]^{2l_3} [C-D]^{2l_4+1}[E-F]^{2h_3}\cdot\nonumber\\ 
  &&[E+F]^{2h_4} \cdot [f_a(t) \cdot f_b(u)].
\end{eqnarray} 
This expression can be rewritten as
\begin{eqnarray}\label{relevexpR2}
{\cal Q}=  &\sum_{{N_iM_i}} a_{{N_iM_i}}\{&A^{N_\mu+N_\nu}B^{M_\mu}(-B)^{M_\nu}
  C^{N_\omega+N_\l}D^{M_\omega}(-D)^{M_\l} \nonumber\\
  &&E^{N_\r+N_\s}(-F)^{M_\r}F^{M_\s}\cdot[f_a(u) \cdot
  f_b(t)]+\nonumber\\
  &+&A^{N_\mu+N_\nu}(-B)^{M_\mu}B^{M_\nu}
  C^{N_\omega+N_\l}(-D)^{M_\omega}D^{M_\l} \nonumber\\
  &&E^{N_\r+N_\s}F^{M_\r}(-F)^{M_\s}\cdot[f_a(t) \cdot
  f_b(u)]\,\,\,\,\, \},
\end{eqnarray}
where $i$ is a compact way to indicate $\m,\n,\omega,\l,\r,\s$, and
where the sum is over 
\begin{eqnarray}
N_\m+M_\m=n_1, &N_\n+M_\n=n_2,&
N_\omega+M_\omega=2k_4+1,\nonumber\\
N_\l+M_\l=2k_3,& N_\r+M_\r=2h_4,&N_\s+M_\s=2h_3.
\end{eqnarray}
We notice that $B$, $C$ and $F$ contain the factor $\cos\pi p$. 
Taking the relation
$$A^a\cdot B^b\cdot C^c\cdot D^d\cdot E^e\cdot F^f=c(\sin\pi
p,\cos\pi\nu_3,\sin\pi\nu_3) \cdot [\cos\pi p]^{b+c+f}
$$
into account,
eq.(\ref{relevexpR2}) can be rewritten as
\begin{eqnarray}\label{relevexpR3}
&&{\cal Q}= \sum_{N_iM_i} a_{N_iM_i} c_{N_iM_i}(\sin\pi p,
  \cos\pi\nu_3,\sin\pi\nu_3) \cdot [\cos\pi
  p]^{M\m+M\n+N\omega+N\l+M\r+M\s} \nonumber\\
  &&\cdot[(-1)^{M\n+M\l+M\r}f_a(u) f_b(t)+(-1)^{M\m+M\omega+M\s}f_a(t)
  f_b(u)]. 
\end{eqnarray}

Now we can show why the quantity (\ref{azzera}) vanishes after the symmetrization
with respect to
$k_1\leftrightarrow k_2$.

\begin{itemize}
  \item If $M_\n+M_\l+M_\r$ has the same parity of $M_\m+M_{\omega}+M_\s$,
  then
  \begin{enumerate}
    \item the \textbf{symmetric} 
    factor $[f_a(u) f_b(t)+f_a(t) f_b(u)]$ appears;
    \item the exponent $M_\m+M_\n+N_{\omega}+N_\l+M_\r+M_\s$ 
    of $\cos\pi p$ is \textbf{odd}, because
          \begin{itemize} 
          \item if $M_\l+M_{\omega}$ is even, then
	  \begin{description}
	   \item \textit{a)} $N_\l+N_\omega$ is odd,
	   \item \textit{b)} $(M_\n+M_\r)+(M_\m+M_\s)$ is even; 
	  \end{description}
	  \item if $M_\l+M_\omega$ is odd, then
	  \begin{description}
	   \item \textit{a)} $N_\l+N_\omega$ is even,
	   \item \textit{b)} $(M_\n+M_\r)+(M_\m+M_\s)$ is odd. 
	  \end{description}
	  \end{itemize}
          \end{enumerate}
  In this case we have that the integral in $p$ is odd under the exchange
  $k_1\leftrightarrow k_2$ and that it is multiplied by a factor even under 
  such an exchange. 
  \item If $M_\n+M_\l+M_\r$ has the opposite parity of $M_\m+M_\omega+M_\s$,
  then
  \begin{enumerate}
    \item the \textbf{antisymmetric} factor 
    $[f_a(u) f_b(t)-f_a(t) f_b(u)]$ appears;
    \item the exponent $M_\m+M_\n+N_\omega+N_\l+M_\r+M_\s$ 
    of $\cos\pi p$ is \textbf{even}, because
          \begin{itemize}
          \item if $M_\l+M_\omega$ is even, then
	  \begin{description}
	   \item \textit{a)} $N_\l+N_\omega$ is odd,
	   \item \textit{b)} $(M_\n+M_\r)+(M_\m+M_\s)$ is odd; 
	  \end{description}
	  %and so $M\m+M\n+N\omega+N\l+M\r+M\s$ is even;
	  \item if $M_\l+M\omega$ is odd, then
	  \begin{description}
	   \item \textit{a)} $N_\l+N_\omega$ is even,
	   \item \textit{b)} $(M_\n+M_\r)+(M_\m+M_\s)$ is even. 
	  \end{description}
	  %and so $M\m+M\n+N\omega+N\l+M\r+M\s$ is even.
	  \end{itemize}
      \end{enumerate}
  In this case we have that the integral in $p$ is even under the exchange
  $k_1\leftrightarrow k_2$ and that it is multiplied by a factor which is odd. 
\end{itemize}

\smallskip

The case in which $\a_4-\a_1$ is even (and consequently $\a_3-a_2$ is odd) can be
treated in the same way.

\section{The open-string threshold}

The amplitude (\ref{ampl24tach2R}) is the sum of three graphs, according to the
integration regions of the variables $\nu_i$'s. In the following we explicitly
discuss the configuration reported in fig.1, namely eq.(\ref{ampl24tach2R}) with
the integrations restricted to the region $1\ge\nu_1\ge \nu_2\ge \nu_3\ge 0$. 

When $l \sim \infty$:
   
\begin{eqnarray}\label{asym}
	&&\eta \left(\frac{2\pi i\alp}{l}\right) \sim \left( 
      	\frac{2\pi \alp}{l}\right)^{-1/2}e^{-l/24\alp}; \\
	&&\vartheta_4\left(\nu,\frac{2\pi i\alp}{l}\right) 
	\sim \left(\frac{2\pi\alp}{l}\right)^{-1/2} 
	e^{\frac{l}{2\alp}(\nu-\nu^2)}e^{-\frac{l}{8\alp}};\\
	&&\vartheta_1\left(\nu,\frac{2\pi i\alp}{l}\right) 
	\sim \left(\frac{2\pi\alp}{l}\right)^{-1/2} 
	e^{\frac{l}{2\alp}(\nu-\nu^2)}e^{-\frac{l}{8\alp}}.
\end{eqnarray}

Hence the asymptotic behaviour of (\ref{ampl24tach2R}) is
\begin{eqnarray} \label{ampl24tach3R} 
  && A^{as}_{S}  =  \mathcal{N}_{\cal S}\mathcal{K}
              \int^\infty dl \,l^{3-d/2}  
	      e^{-\frac{Y^2l}{2\pi\alp}}
	      e^{-\frac{i}{2}k_1\th k_2}
	      e^{\frac{i}{2}k_3\th k_4} \\
	  & \times  & \left( \prod_{n=1}^3 \int_0^{\nu_{n-1}}d\nu_n \right)
	      e^{i(\nu_{12})k_1\th k_2}
	      e^{-i\nu_3k_3\th k_4}e^{l\sum_{n<m=1}^{4}\nu_{nm}(1-\nu_{nm})
    	      k_nG^{-1}k_m}\nonumber .
\end{eqnarray}
We change the integration variables, introducing 
\begin{eqnarray}\label{chan}
  x_1 &=& \nu_1-\nu_2 \nonumber\\
  x_2 &=& \nu_2-\nu_3 \nonumber\\
  x_3 &=& \nu_3       \nonumber\\
  x_4 &=& 1-\nu_1,     
\end{eqnarray}
such that $x_1+x_2+x_3+x_4=1$. In these new variables we get:
\begin{eqnarray} \label{sumNUR}
&&  \sum_{n<m}^4\nu_{nm}(1-\nu_{nm})k_nG^{-1}k_m =
  -2k_1G^{-1}k_2(x_2+x_3)x_4 -2k_1G^{-1}k_3x_3x_4 \nonumber\\
  &&-2k_2G^{-1}k_3(x_1+x_4)x_3 - k_1G^{-1}k_1(x_1+x_2+x_3)x_4 \nonumber\\
  &&-k_2G^{-1}k_2(x_1+x_3)(x_2+x_4) -k_3G^{-1}k_3(x_1+x_2+x_4)x_3.
\end{eqnarray}
We impose the on-shell condition on the external momenta 
\begin{equation} \label{onshellR}
  k_nG^{-1}k_n=0
\end{equation}
and introduce the Mandelstam invariants
\begin{eqnarray} \label{MandelR}
  s&=&-(k_1+k_2)G^{-1}(k_1+k_2),  \nonumber\\  
  t&=&-(k_1+k_4)G^{-1}(k_1+k_4),  \nonumber\\
  u&=&-(k_1+k_3)G^{-1}(k_1+k_3). 
\end{eqnarray}
Considering these relations, eq.(\ref{sumNUR}) becomes 
\begin{equation}
  \sum_{n<m}^4\nu_{nm}(1-\nu_{nm})k_nG^{-1}k_m = s(x_2+x_3)x_4 +ux_3x_4 + t(x_1+x_4)x_3 
\end{equation}
and the amplitude (\ref{ampl24tach3R}) can be rewritten as
\begin{eqnarray} \label{ampl24tach4R} 
  A^{as}_{S} & = & \mathcal{N}_{\cal S}\mathcal{K}
              \int^\infty dl \,l^{3-d/2}  
	      e^{-\frac{i}{2}k_1\th k_2}
	      e^{\frac{i}{2}k_3\th k_4} \left( \prod_{n=1}^4 \int_0^1 dx_n \right)
	      \delta(1-\sum_n x_n)\nonumber \\
	  & \times  & 
	      e^{i(\nu_{12})k_1\th k_2}
	      e^{-i\nu_3k_3\th k_4}e^{-l(m^2 - s(x_2+x_3)x_4 - ux_3x_4 - t(x_1+x_4)x_3)},
\end{eqnarray}
where we have defined $m^2\equiv\frac{Y^2}{(2\pi\alp)^2}$.
The function $A^{as}_{S}(s)$ is well defined when the integral converges, 
namely when 
\begin{eqnarray}
  m^2 - s(x_2+x_3)x_4 - ux_3x_4 - t(x_1+x_4)x_3\,>\, 0 && \  \forall {x_1,x_2,x_3,x_4}.
\end{eqnarray}
This condition requires
\begin{eqnarray}
  s&<&\min_{\{x\}} \left[ \frac{m^2}{(x_2+x_3)x_4} + (-u)\frac{x_3}{(x_2+x_3)} + 
	(-t)\frac{(x_1+x_4)x_3}{(x_2+x_3)x_4}\right] 
	=\nonumber\\
  &=& m^2\, \min_{0<x_2<1}\left[\frac{1}{x_2(1-x_2)}  \right] =  4m^2.
\end{eqnarray}
We have thereby recovered the expected threshold condition. Analogous contribution 
is obtained by completing the integration region of the variable $\nu_3$.

\end{document}